\documentclass[preprint,superscriptaddress]{revtex4}
 \pdfoutput=1 

\usepackage{amsmath}    
\usepackage{graphicx}   
\usepackage{color}
\usepackage[10pt]{moresize}

\usepackage{graphics,graphicx}
\usepackage{amsfonts}
\usepackage{amssymb}
\usepackage{amscd}
\usepackage{amsmath}
\usepackage{enumerate}
\usepackage{epsfig}
\usepackage{subfigure}
\usepackage[perpage,symbol]{footmisc}

\usepackage{booktabs}

\newcommand{\ket}[1]{\mbox{$\left| #1 \right\rangle$}}

\begin{document}

\title{Ground test of satellite constellation based quantum communication}

\author{Sheng-Kai Liao$^{\footnotemark[1]}$}
\author{Hai-Lin Yong$^{\footnotemark[1]}$}
\author{Chang Liu$^{\footnotemark[1]}$}
\author{Guo-Liang Shentu$^{\footnotemark[1]}$}
\author{Dong-Dong Li}
\author{Jin Lin}
\author{Hui Dai}
\affiliation {Shanghai Branch, National Laboratory for Physical Sciences at Microscale and Department of Modern Physics, University of Science and Technology
of China, Shanghai, 201315, China.}
\affiliation{Synergetic Innovation Center of Quantum Information and Quantum Physics, University of Science and Technology of China, Shanghai, 201315, China}
\author{Shuang-Qiang Zhao}
\affiliation{School of Information Science and Engineering, Ningbo University, Ningbo 315211, China.}
\author{Bo Li}
\author{Jian-Yu Guan}
\author{Wei Chen}
\author{Yun-Hong Gong}
\author{Yang Li}
\affiliation {Shanghai Branch, National Laboratory for Physical Sciences at Microscale and Department of Modern Physics, University of Science and Technology
of China, Shanghai, 201315, China.}
\affiliation{Synergetic Innovation Center of Quantum Information and Quantum Physics, University of Science and Technology of China, Shanghai, 201315, China}
\author{Ze-Hong Lin}
\affiliation{School of Information Science and Engineering, Ningbo University, Ningbo 315211, China.}
\author{Ge-Sheng Pan}
\affiliation {Shanghai Branch, National Laboratory for Physical Sciences at Microscale and Department of Modern Physics, University of Science and Technology
of China, Shanghai, 201315, China.}
\affiliation{Synergetic Innovation Center of Quantum Information and Quantum Physics, University of Science and Technology of China, Shanghai, 201315, China}
\author{Jason S. Pelc}
\author{M. M. Fejer}
\affiliation{Edward L. Ginzton Laboratory, Stanford University, Stanford, California 94305, USA.}
\author{Wen-Zhuo Zhang}
\affiliation {Shanghai Branch, National Laboratory for Physical Sciences at Microscale and Department of Modern Physics, University of Science and Technology of China, Shanghai, 201315, China.}
\affiliation{Synergetic Innovation Center of Quantum Information and Quantum Physics, University of Science and Technology of China, Shanghai, 201315, China}
\author{Wei-Yue Liu}
\affiliation{School of Information Science and Engineering, Ningbo University, Ningbo 315211, China.}
\author{Juan Yin}
\author{Ji-Gang Ren}
\affiliation {Shanghai Branch, National Laboratory for Physical Sciences at Microscale and Department of Modern Physics, University of Science and Technology
of China, Shanghai, 201315, China.}
\affiliation{Synergetic Innovation Center of Quantum Information and Quantum Physics, University of Science and Technology of China, Shanghai, 201315, China}
\author{Xiang-Bin Wang}
\affiliation{Synergetic Innovation Center of Quantum Information and Quantum Physics, University of Science and Technology of China, Shanghai, 201315, China}
\affiliation{Jinan Institute of Quantum Technology, Shandong Academy of Information and Communication Technology, Jinan 250101, China.}
\author{Qiang Zhang}
\affiliation {Shanghai Branch, National Laboratory for Physical Sciences at Microscale and Department of Modern Physics, University of Science and Technology
of China, Shanghai, 201315, China.}
\affiliation{Synergetic Innovation Center of Quantum Information and Quantum Physics, University of Science and Technology of China, Shanghai, 201315, China}
\affiliation{Jinan Institute of Quantum Technology, Shandong Academy of Information and Communication Technology, Jinan 250101, China.}
\author{Cheng-Zhi Peng}
\author{Jian-Wei Pan}
\affiliation {Shanghai Branch, National Laboratory for Physical Sciences at Microscale and Department of Modern Physics, University of Science and Technology
of China, Shanghai, 201315, China.}
\affiliation{Synergetic Innovation Center of Quantum Information and Quantum Physics, University of Science and Technology of China, Shanghai, 201315, China}

\footnotetext[1]{These authors contributed equally to this work.}


\begin{abstract}
Satellite based quantum communication has been proven as a feasible way to achieve global scale quantum communication network\cite{Peng:13km:2005, Zeilinger:Decoy:2007,Jin:Teleportation:2010,Yin:Teleportation:2012,XSMa:Teleportation:2012,Wang:Direct:2013,Nauerth:AirQKD:2013,Yin:single:2013,Vallone:Single:2015}.
Very recently, a low-Earth-orbit (LEO) satellite has been launched\cite{Xin:Chinese:2011} for this purpose.
However, with a single satellite, it takes an inefficient 3-day period\cite{Fritz:Revisit:1996} to provide the worldwide connectivity. On the other hand, similar to how the Iridium system\cite{Pratt:IRIDIUM:1999} functions in classic communication, satellite constellation (SC) composed of many quantum satellites, could provide global real-time quantum communication. In such a SC, most of the satellites will work in sunlight. Unfortunately, none of previous ground testing experiments\cite{Peng:13km:2005, Zeilinger:Decoy:2007,Jin:Teleportation:2010,Yin:Teleportation:2012,XSMa:Teleportation:2012,Wang:Direct:2013,Nauerth:AirQKD:2013,Yin:single:2013,Vallone:Single:2015} could be implemented at daytime. During daytime, the bright sunlight background prohibits quantum communication in transmission over long distances. In this letter, by choosing a working wavelength of 1550 nm and developing free-space single-mode fibre coupling technology and ultra-low noise up-conversion single photon detectors\cite{Shentu:Detector:2013}, we overcome the noise due to sunlight and demonstrate a 53-km free space quantum key distribution (QKD) in the daytime through a 48-dB loss channel. Our system not only shows the feasibility of satellite based quantum communication in daylight, but also has the ability to naturally adapt to ground fibre optics, representing an essential step towards a SC-based global quantum network.

\end{abstract}

\maketitle

\begin{figure}[tbh]
\centering
\resizebox{16cm}{!}{\includegraphics{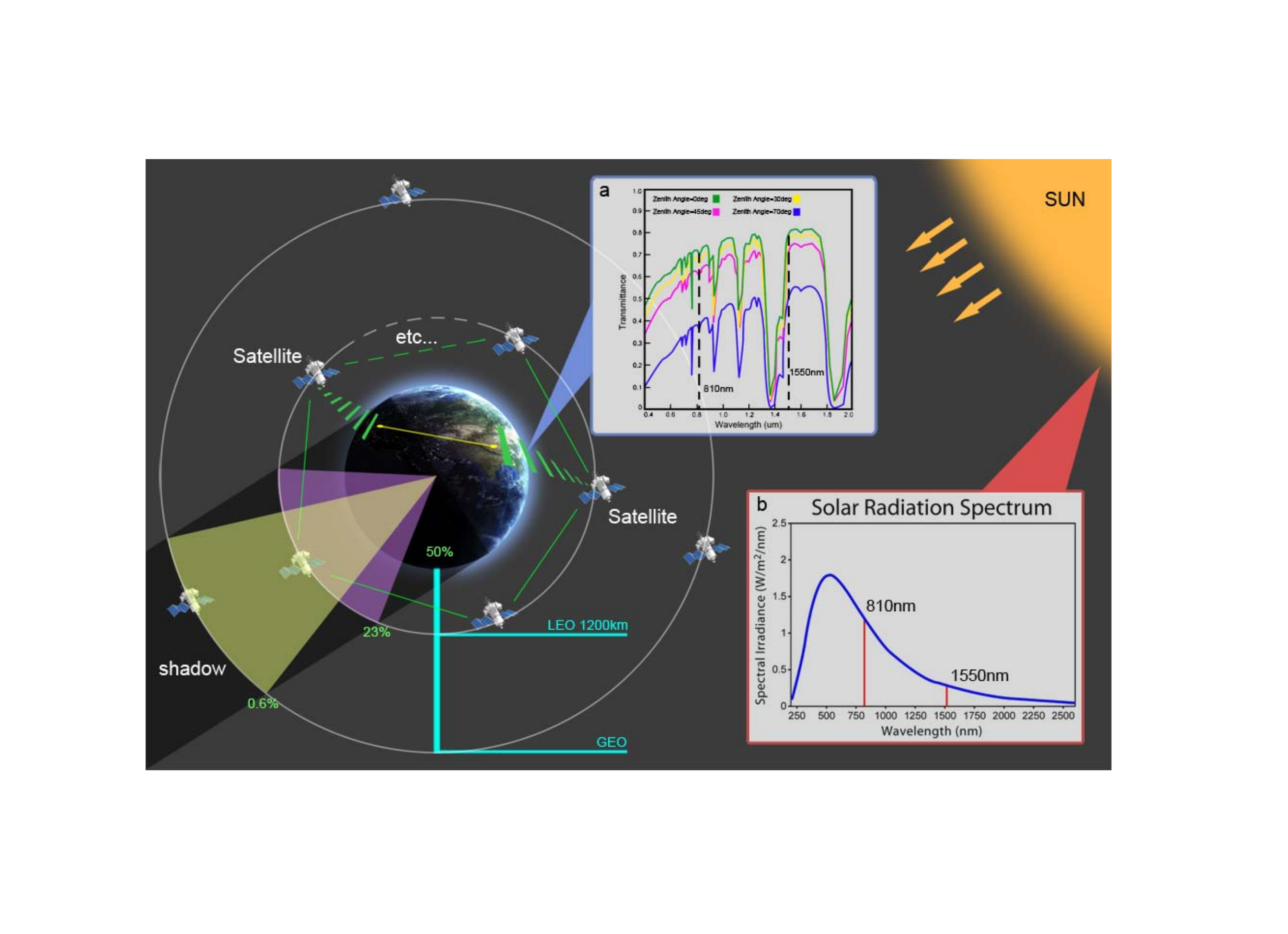}}
\caption{\textbf{Satellite-constellation based global quantum network.} A global quantum network needs many LEO satellites or several geosynchronous orbit satellites to compose a satellite constellation (SC). The time of a satellite in the Earth shadow area, which we call it night, is inverse proportional to the orbit height of the satellite;
\textbf{a}) Transmittance spectra from visible to near infrared light in atmosphere at selected zenith angles; \textbf{b}) Solar radiation spectrum from visible to near infrared light.}
\label{Fig:Daylight:SC}
\end{figure}

A SC is expected to operate in LEO satellites, or high-Earth-orbit satellites such as geosynchronous orbit (GEO) satellites. The probability of a satellite being in the Earth shadow zone drops rapidly with increasing orbit height (Fig~\ref{Fig:Daylight:SC}). A LEO satellite system has a ~70\% probability being in the sunlight area, and for GEO satellite, this probability rise to ~99\% in the sunlight area\cite{Gilmore:spacecraft_thermal:2002}. Meanwhile, the total channel loss between a LEO satellite and the Earth and between LEO satellites is typically around 40-45 dB\cite{Pfennigbauer:Intersatellite:2003,Tomaello:intersatellite:2011}. Therefore, in order to test the feasibility of an SC based quantum network, quantum communication through at least a $\geq$40-dB loss channel in daylight case is critically required.

There have been several pioneering experiments on daylight quantum communication prior to our work\cite{Buttler:Daylight:2000,Hughes2002:10km,Hockel:Ultranarrow:2009,Restelli:Timing:Daylight:2010,Shan:Rbvapor:Daylight:2006,Rogers:Halpha:Daylight:2006,Peloso:Entanglement:Daylight:2009}. Although the experiments were novel, the maximum loss calculated from them was only ~20 dB. The main cause of the unsatisfactory performance was the strong background noise from the scattered sunlight, which was typically 5 orders of magnitude greater than the background noise during the night time\cite{Miao:QKDNoise:2005}. To reduce this noise, we first switched the working wavelength to 1550.14 nm from the 700-800 nm used in all previous experiments.

1550 nm is also known to be an atmospheric window. In fact, the transmission efficiency is slightly higher at 1550 nm than at 800 nm as shown in Fig~\ref{Fig:Daylight:SC} a, and from the solar spectrum in Fig~\ref{Fig:Daylight:SC} b, we can see that the sunlight intensity at 1550 nm is around 5 times weaker than it is at 800 nm. Furthermore, the main type of scattering for links either between a satellite and the Earth or between two satellites is Rayleigh scattering, whose intensity is proportional to $1/\lambda^{4}$. Therefore, Rayleigh scattering at 1550 nm is only 7\% of its value at ~800 nm. In total, the background noise with 1550 nm light can be reduced to 3\% of the background noise with 800 nm light. We measured the noise count rate of 1550 nm light in the daylight case by pointing a telescope at the sky to simulate satellite-to-Earth communication. The result is was smaller by a factor of 22.5\footnotemark[2]\footnotetext[2]{Note that without a satellite, all existing free-space experiments, including this work, were implemented on the Earth, where the direction for the free-space communication is parallels to the Earth, rather than pointing at the sky. In this condition, Mie scattering, which does not follow the $1/\lambda^{4}$ relation, will be the main noise source instead of Rayleigh scattering.} to the case of 850 nm light. Moreover, 1550 nm is the telecom-band wavelength and is widely used for fibre-optical communication. Using the same wavelength for both free-space and fibre-optical communication is an optimal choice.

Despite the advantages of 1550 nm, researchers have been reluctant to use this wavelength due to lack of good commercial single-photon detectors for the telecom-band. Here we have developed a compact
up-conversion single-photon detector (SPD)\cite{Shentu:Detector:2013} (Fig~\ref{Fig:Daylight:Setup}e). In the up-conversion detector, a telecom-band photon is mixed with a strong pumping signal of 1950 nm in a wavelength division multiplexing (WDM) coupler and then sent to a fibre-pigtailed periodically poled lithium niobate (PPLN) waveguide. The generated photons are collected by an anti-reflectively-coated objective lens, and are then separated from the pump and the spurious light by a dichroic mirror, a short pass filter and a bandpass filter. A volume Bragg grating(VBG) with 95\% reflection efficiency is exploited to further suppress the noise from both spontaneous Raman scattering generated in the nonlinear process and the sunlight background. In our setup, the VBG filter's spectral bandwidth is 0.05 nm at a centre wavelength tunable near 864 nm (full width at half maximum), corresponding to a bandwidth of 0.16 nm at the signal wavelength. Finally, the SFG photons are collected and detected by a Silicon APD. Using a pump power of 200 mW, the total-system detection efficiency is 8\%, with an average dark count rate of approximately 20 Hz, which is the same as the intrinsic dark count of the Si APD. Note that, in all previous daylight experiments, narrow band filters, including Fabry-Perot cavities and atomic vapours\cite{Hockel:Ultranarrow:2009,Shan:Rbvapor:Daylight:2006} were exploited to reduce the noise. For our experiment, the phase matching condition and VBG of the detector itself realize spectrum filtering, and no additional filters are needed.

To further improve the signal-to-noise ratio, we developed single-mode fibre coupling technology for spatial filtering. We reduce the receiving aperture by single-mode fibre coupling to improve the signal-to-noise ratio, making use of the fact that the signal light propagates in certain direction while the scattered sunlight is isotropic (Fig~\ref{Fig:Daylight:Setup} c). The receiving angle of our system was ~6 $\mu$rad, with a whose corresponding receiving field that was two orders of magnitude smaller than in previous experiments\cite{Zeilinger:Decoy:2007,Yin:Teleportation:2012,XSMa:Teleportation:2012}. Compared with previous free space experiments that used many optical elements\cite{Wang:Direct:2013,Nauerth:AirQKD:2013}, we used the least amount of optics to design the receiving telescope and to reduce the optical attenuation and abbreviation. The focus length of the off-axis primary mirror was set to 2000 mm in order to optimize the single-mode fibre coupling efficiency. We put only one fast-steering mirror (FSM) and one dichroic mirror (DM) between the primary mirror and the single-mode fibre, with which we also developed an optical tracking system (see method) with 300Hz feedback frequency to stabilize the single-mode fibre coupling. With this setup, we could obtain a single-mode fibre coupling efficiency of over 30\% in the lab (with $0.08\mu rad$ tracking precision). While in the outdoor experiment, the single-mode fibre coupling efficiency was reduced to 5\% by horizontal air turbulence (with $3\mu rad$ tracking precision). Here, we would like to emphasize that 5\% efficiency is still much higher than the 0.1\% efficiency in previous experiments\cite{Ren:Teleportation:2009}. Furthermore, in future ground-to-satellite experiments, we will receive signals vertically and the turbulence is avoided. A efficiency of 30\%, similar to that in the lab test, can be expected.

Fig~\ref{Fig:Daylight:Setup} shows the setup of our experiment on Qinghai lake. The sending terminal (Alice) was located at Heimahe village $(N37^{\circ}03^{'}35.9^{''}, E99^{\circ}47^{'}20.9^{''})$, Qinghai Province, China, as shown in Fig~\ref{Fig:Daylight:Setup} a. It sent the 1550 nm signal beam through a 53-km-free-space link across Qinghai Lake to the receiving terminal (Bob), which was located at Quanji village $(N37^{\circ}16^{'}42.5^{''}, E99^{\circ}52^{'}59.9^{''})$, Qinghai Province.

At the sending terminal, we developed a 1550-nm light source with a decoy scheme\cite{Hwang:Decoy:2003,Wang:Decoy:2005,Lo:Decoy:2005}. We used four DFB lasers with a central wavelength at 1550.14 nm and full-width-at-half-maximum (FWHM) of 0.02 nm to emit 500 ps pulses. We created a standard BB84 source by combining the pulses with two polarizing beam splitters (PBS), one beam splitter (BS) and one variable optical attenuator. At a clock frequency of 100 MHz, the source randomly generate one of the four polarization states, $|H\rangle$, $|V\rangle$, $|+\rangle$, $|-\rangle$ with one of three average photon numbers per pulse, ($0.6,0.14,0$). Here $|H\rangle/|V\rangle$ represents horizontal/vertical polarization, $|+\rangle = (|H\rangle+|V\rangle)/\sqrt{2}$ and $|-\rangle = (|H\rangle-|V\rangle)/\sqrt{2}$. The intensity of the signal state was 0.6 per pulse. We used two decoy states, one was the vacuum state and the other was the state with average photon number 0.14. The probability ratio of the one signal and the two decoy states is 2:1:1. All random control signals were generated at a high-speed by a random number generator. After that, the signal was coupled into the single-mode fibre and collimated out into free-space with a triplet collimator (TC), and then sent to the telescope that was mounted on a 2-dimensional platform (Fig~\ref{Fig:Daylight:Setup} b). We minimized the sending divergence angle of the 1550 nm beam to reduce geometric loss. We chose a Schmidt-Cassegrain type system with a d=254 mm primary mirror for sending as the divergence angle is proportional to the wavelength and inversely proportional to the primary mirror's size. The measured divergence angle was $12~\mu rad$ which is very close to the diffraction limit and four times smaller than our previous sending system\cite{Wang:Direct:2013}.

At the receiving terminal, the signal light was collected by a receiving telescope as shown in Fig~\ref{Fig:Daylight:Setup} c, which consisted of a primary parabolic mirror with a diameter of 420 mm and focal length of 2000 mm and a single-mode fibre coupling module with an optical tracking system. The signal photons went through a 20-m long single-mode fibre to the detection system. A fibre BS is used to select a measurement basis, and two fibre PBSs together with four up-conversion SPDs were used for detection (Fig~\ref{Fig:Daylight:Setup} d). All detected signals were to be fed into a time-to-digital converter (TDC) for analysis. The time window was set to 1 ns, for timely filtering of the noise as in previous experiments\cite{Hughes2002:10km,Restelli:Timing:Daylight:2010}. We also developed an efficient self-synchronized system based only on GPS, which could reduce the noise and attenuation better than previous pulsed-laser synchronization systems\cite{Yin:Teleportation:2012}.

In our QKD experiment, Alice and Bob extracted the final secure key out of the raw data following a standard decoy BB84 post-processing procedure\cite{Wang:Decoy:2005,Lo:Decoy:2005,Fung:Finite:2010}. The final key rate formula is
\begin{equation} \label{MIExp:Post:KeyrateMI}
\begin{aligned}
R_{pulse} &\geq q p_\mu \{-Q_\mu f(E_\mu)H_2(E_\mu)+Q_1[1-H_2(e_1)]\},
\end{aligned}
\end{equation}
where $q=1/2$ is the basis reconciliation factor, $p_\mu$ is the probability for emitting signal states, $Q_\mu$ and $E_\mu$ are the gain and the error rate of the signal states, respectively and $f$ is the error correction efficiency. The low-density parity-check (LDPC) code was used for error correction, with $H_2(e)=-e\log_{2}(e)-(1-e)\log_{2}(1-e)$ being the binary Shannon entropy function, and $Q_1$ ($e_{1}$) being the gain (phase error rate) when source generate single-photon states. We run our QKD system for 464 seconds. The results are listed in Table \ref{Tab:Daylight:Key}. The final key rate was 20-400 bits per second. The variation of the final key rate was mainly due to the channel loss changes in the atmospheric environment. We did and also repeated QKD experiments from 15:30 to 16:30 at local time for several sunny days. The total key size was larger than $10^5$ bits in 10 minutes.

We remark that the internal modulation of the decoy/signal states guaranteed our system security against photon number splitting\cite{HIGM:PNS:1995,BLMS:PNS:2000} and unambiguous-state-discrimination attack\cite{LoPreskill:NonRan:2007}. The total loss over our 53-km-free-space QKD was 48 dB, which consisted of a 14 dB single-mode fibre coupling loss and 34 dB channel loss (including geometric loss, air attenuation, receiving loss, and detection loss). Our experiment showed the success of QKD through a 48 dB loss channel, which offers strong support for an SC based quantum network.


In summary, we successfully demonstrated a 53-km free space QKD in daylight. With a working wavelength of 1550 nm, up-conversion SPD and single-mode fibre coupling, we offer a solution to the problem of quantum communication in daylight. Our work proves the feasibility of a LEO quantum SC which works mostly in the daylight. Moreover, our work also offers an optimal option for a global quantum network consisting of a quantum SC and the existing ground fibre networks. Our free space single mode coupling technology is also very useful for both free-space quantum communication and laser communication applications such as measurement-device-independent QKD\cite{Lo:MIQKD:2012}, quantum teleportation\cite{BBCJPW_93}, quantum repeaters\cite{Briegel1998:repeater}, quantum metrology\cite{Giovannetti:QEM:2004}, homodyne coherent detection\cite{Yurke:Homodyne:1985} in laser ranging. Moreover, the SC based quantum network can also find rapid application in precisely sharing timing information globally\cite{Komar:clock:2014}.

The communication distance and secure key rate still have room to improve.
The system's repetition rate can be increased by increasing the up-conversion detector rate,
which may become as large as 2 GHz\cite{Gisin_up2_06}. With cavity-enhanced Si APD,
we can increase the efficiency of the up-conversion detector to improve the secure key rate.
Meanwhile, for future satellite-to-ground communication with a very small zenith angle,
Rayleigh scattering will dominate, and we could have an even better signal-to-noise ratio than i
n our ground experiment with 1550-nm light. 

\section*{Acknowledgments}
The authors would like to thank Yu-Ao Chen, Yuan Cao, Yang Liu, Yu Xu for enlightening discussions.
This work has been supported by the National Fundamental Research Program (under Grant No. 2013CB336800), the ``Strategic Priority Research Program'' of the Chinese Academy of Sciences (under Grant No. XDA04020000), the NNSF of China, the Chinese Academy of Sciences, and the 10000-Plan of Shandong province (TaiShan Scholars).

\section*{Competing financial interests}
The authors declare no competing financial interests.

\section*{METHODS SUMMARY}

\textbf{Optical tracking, alignment, and calibration.}

\begin{figure}[tbh]
\centering
\resizebox{16cm}{!}{\includegraphics{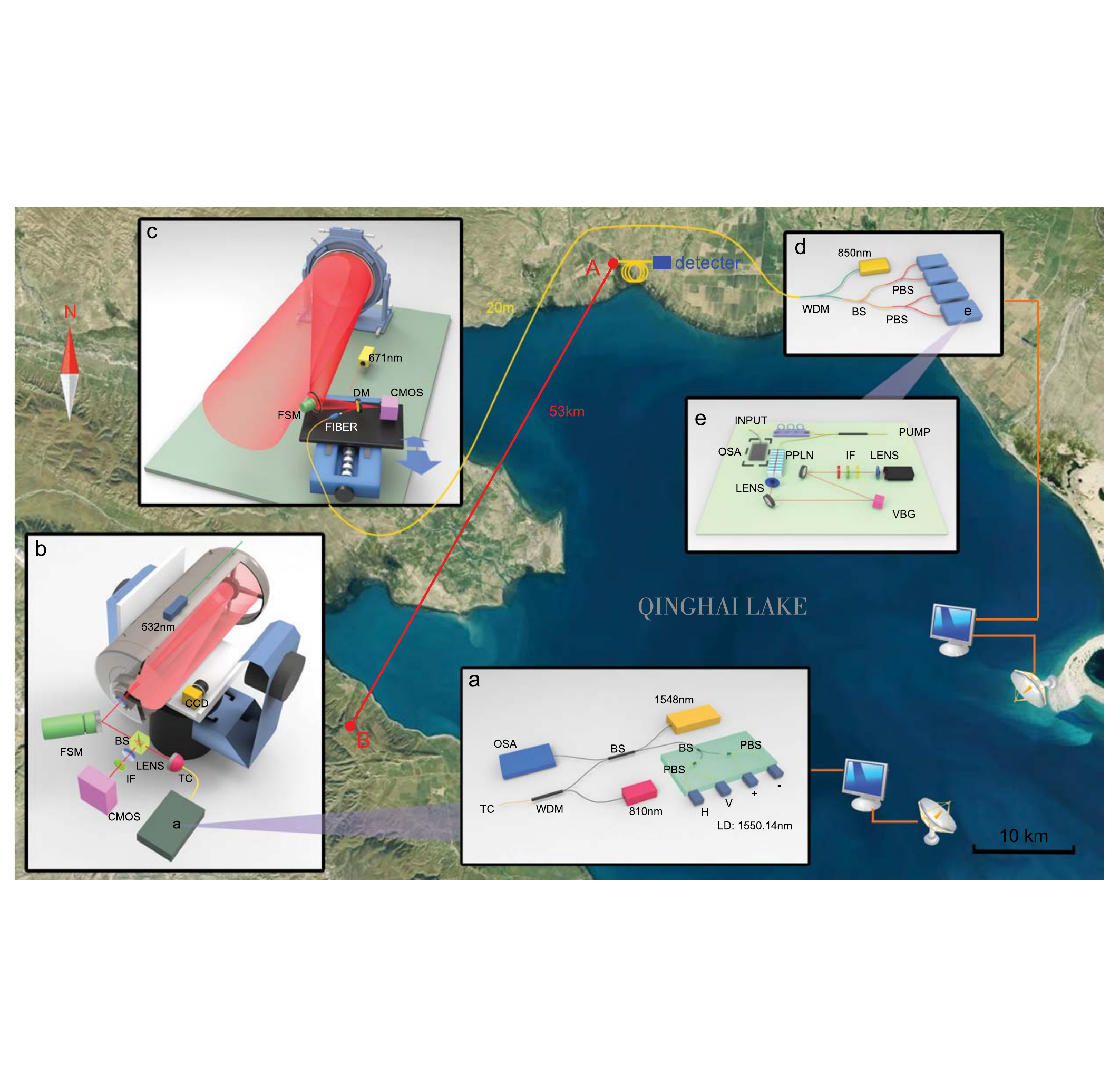}}
\caption{\textbf{Bird¡¯s-eye view for the 53-km QKD experiment in daylight.} Alice and Bob each locates at one side of the Qinghai Lake.
\textbf{a}) The 1550 nm lasers are encoded to four quantum states ($\ket{H},\ket{V},\ket{+},\ket{-}$) by 2 PBSs and 1 BS. A 810 nm and a 1548 nm beacon lasers are combined and sent to triplet collimator (TC) for optical alignment and tracking. An optical spectrum analyser (OSA) is used to calibrate the signal spectrum;
\textbf{b}) The sending terminal has a telescope system on a 2-axial rotation stage and an optical tracking system; 
\textbf{c}) The receiving terminal has an off-axial parabolic mirror and a single-mode fibre coupling module; 
\textbf{d}) Received photons are transmitted to the detector via a 20 m fibre. At the detection part, photons are measured with 2 PBSs, 1 BS and 4 detectors;
\textbf{e}) Up-conversion single-photon detector modules. Narrow bandwidth VBGs are used to narrow the working spectrum of the detectors and reduce the noise;
For the details of optical alignment and tracking, see Method.}
\label{Fig:Daylight:Setup}
\end{figure}

First, a 500 mW, 532 nm beacon laser, coaxially equipped on the sending telescope, is pointed to the receiving site. We manually align the parabolic mirror to make sure that the green laser spot locates at the mirror's center. Then a 2 W, 671 nm beacon laser, coaxially equipped with the receiving telescope, is pointed to the sending site. In the sending sites, a wide-field camera is installed behind the guide scope to take photos of the red beacon light. The photo information is fed to the two dimensional sending platform, which shall align the telescope's direction to achieve an optimal tracking of the 671 nm beacon light. This constitutes the coarse tracking system in sending site. The fine tracking system consists of a beam splitter (BS), a fast steering mirror (FSM), a interference filter (IF), and a complementary metal oxide semiconductor (CMOS) imaging sensor, as is shown in Fig~\ref{Fig:Daylight:Setup}. The BS is used to collect the 671 nm beacon light to fine tracking CMOS sensor, while FSM is used to fine adjust optical path according to a correction program through the image information obtained by the fine tracking CMOS imaging sensor.

In the sending site, except for the 532 nm laser, we also equipped two additional beacon light. One is 10 mW, 810 nm laser and the other is 1 W, 1548 nm. Both lasers are mixed with the signal light with a WDM, and a BS, respectively, as shown in Fig~\ref{Fig:Daylight:Setup}. The two lasers are utilized as beacon lights for aligning and tracking single mode fibre coupling. In the receiving site, the single mode fibre coupling system, mounted on a one dimension translation stage, consists of a FSM, a DM and a CMOS. The FSM is just the secondary mirror of the receiving telescope. And the DM mirror reflects the telecomband light and transmits the 810 nm beacon light. The reflected telecomband light is collimated into a single mode fibre and the transmitted 810 nm light is captured by the CMOS imaging sensor. Before the QKD experiment, we combine a 810 nm and a 1550 nm laser into a 3-meter free space collimator and shine the light to the receiving telescope to simulate the beacon light. With the simulated light, we first put the single mode fibre in the focal point of the parabolic reflector by adjusting the translation stage, then we align the FSM and the DM to optimize fibre coupling efficiency. Once the maximum efficiency is achieved, we record and mark the 810 nm light position by the CMOS sensor. During the QKD experiment, we first make sure that the 810 nm beacon light locates at the marked position of the CMOS imaging sensor. Then, we optimize the translation stage, the FSM, the DM mirror by measuring the single mode fibre's coupled optical power with an InGaAs power meter. Meanwhile, a 850 nm beacon laser is also coupled into the fibre in the other side, propagates oppositely to the 810 nm beacon laser through the free space channel, and captured by the CMOS imaging sensor in the sending site to double check the fine tracking.

In our experiment, the shift of the bandwidth of our lasers and detectors should be guaranteed. We use each high-precision optical spectrum analyzer (OSA) at the source and the detector. The OSA at Fig~\ref{Fig:Daylight:Setup}a is for source, another one is put at the input of Fig~\ref{Fig:Daylight:Setup}e. OSAs are self-calibrated in advance. After the link is established, a strong light (about 1 W optical power) at 1548 nm is sent from the transmitter to the receiver. With OSAs, the wavelength is precisely measured at the two terminals. After all, the wavelength for the source and detector is calibrated with a precision smaller than 0.01 nm at distance.




\clearpage

\begin{table}
\centering
\caption{\textbf{Experimental parameters and results.} $T$ is the effective time for QKD, $Q_\mu$ and $Q_\nu$ are the gain for the signal states and the decoy states, respectively; $Y_0$ is the yield for the vacuum states, $E_\mu$ is the QBER of the signal states, $R_{pulse}$ is final key rate per clock cycle, and $R_{total}$ is the total final key size of the experiment.}
\label{Tab:Daylight:Key}
\begin{tabular}{ccccccccccccc}
\hline\hline
$T$&$Q_\mu$&$Q_\nu$&$Y_0$&$E_\mu$&$E_\nu$&$R_{pulse}$&$R_{total}$\\
\hline
$464s$&$1.63\times10^{-5}$&$4.11\times10^{-6}$&$2.38\times10^{-7}$&$	1.65\%$&$	 3.35\%$&$2.97\times10^{-6}$&$68912~bits$\\
\hline\hline
\end{tabular}
\end{table}

\clearpage

\clearpage


\begin{thebibliography}{39}
\expandafter\ifx\csname natexlab\endcsname\relax\def\natexlab#1{#1}\fi
\expandafter\ifx\csname bibnamefont\endcsname\relax
  \def\bibnamefont#1{#1}\fi
\expandafter\ifx\csname bibfnamefont\endcsname\relax
  \def\bibfnamefont#1{#1}\fi
\expandafter\ifx\csname citenamefont\endcsname\relax
  \def\citenamefont#1{#1}\fi
\expandafter\ifx\csname url\endcsname\relax
  \def\url#1{\texttt{#1}}\fi
\expandafter\ifx\csname urlprefix\endcsname\relax\def\urlprefix{URL }\fi
\providecommand{\bibinfo}[2]{#2}
\providecommand{\eprint}[2][]{\url{#2}}

\bibitem[{\citenamefont{Peng et~al.}(2005)\citenamefont{Peng, Yang, Bao, Zhang,
  Jin, Feng, Yang, Yang, Yin, Zhang et~al.}}]{Peng:13km:2005}
\bibinfo{author}{\bibfnamefont{C.-Z.} \bibnamefont{Peng}},
  \bibinfo{author}{\bibfnamefont{T.}~\bibnamefont{Yang}},
  \bibinfo{author}{\bibfnamefont{X.-H.} \bibnamefont{Bao}},
  \bibinfo{author}{\bibfnamefont{J.}~\bibnamefont{Zhang}},
  \bibinfo{author}{\bibfnamefont{X.-M.} \bibnamefont{Jin}},
  \bibinfo{author}{\bibfnamefont{F.-Y.} \bibnamefont{Feng}},
  \bibinfo{author}{\bibfnamefont{B.}~\bibnamefont{Yang}},
  \bibinfo{author}{\bibfnamefont{J.}~\bibnamefont{Yang}},
  \bibinfo{author}{\bibfnamefont{J.}~\bibnamefont{Yin}},
  \bibinfo{author}{\bibfnamefont{Q.}~\bibnamefont{Zhang}},
  \bibnamefont{et~al.}, \bibinfo{journal}{Phys. Rev. Lett.}
  \textbf{\bibinfo{volume}{94}}, \bibinfo{pages}{150501}
  (\bibinfo{year}{2005}).

\bibitem[{\citenamefont{Schmitt-Manderbach
  et~al.}(2007)\citenamefont{Schmitt-Manderbach, Weier, F\"urst, Ursin,
  Tiefenbacher, Scheidl, Perdigues, Sodnik, Kurtsiefer, Rarity
  et~al.}}]{Zeilinger:Decoy:2007}
\bibinfo{author}{\bibfnamefont{T.}~\bibnamefont{Schmitt-Manderbach}},
  \bibinfo{author}{\bibfnamefont{H.}~\bibnamefont{Weier}},
  \bibinfo{author}{\bibfnamefont{M.}~\bibnamefont{F\"urst}},
  \bibinfo{author}{\bibfnamefont{R.}~\bibnamefont{Ursin}},
  \bibinfo{author}{\bibfnamefont{F.}~\bibnamefont{Tiefenbacher}},
  \bibinfo{author}{\bibfnamefont{T.}~\bibnamefont{Scheidl}},
  \bibinfo{author}{\bibfnamefont{J.}~\bibnamefont{Perdigues}},
  \bibinfo{author}{\bibfnamefont{Z.}~\bibnamefont{Sodnik}},
  \bibinfo{author}{\bibfnamefont{C.}~\bibnamefont{Kurtsiefer}},
  \bibinfo{author}{\bibfnamefont{J.~G.} \bibnamefont{Rarity}},
  \bibnamefont{et~al.}, \bibinfo{journal}{Phys.~Rev.~Lett.~}
  \textbf{\bibinfo{volume}{98}}, \bibinfo{pages}{010504}
  (\bibinfo{year}{2007}).

\bibitem[{\citenamefont{{Jin} et~al.}(2010)\citenamefont{{Jin}, {Ren}, {Yang},
  {Yi}, {Zhou}, {Xu}, {Wang}, {Yang}, {Hu}, {Jiang}
  et~al.}}]{Jin:Teleportation:2010}
\bibinfo{author}{\bibfnamefont{X.-M.} \bibnamefont{{Jin}}},
  \bibinfo{author}{\bibfnamefont{J.-G.} \bibnamefont{{Ren}}},
  \bibinfo{author}{\bibfnamefont{B.}~\bibnamefont{{Yang}}},
  \bibinfo{author}{\bibfnamefont{Z.-H.} \bibnamefont{{Yi}}},
  \bibinfo{author}{\bibfnamefont{F.}~\bibnamefont{{Zhou}}},
  \bibinfo{author}{\bibfnamefont{X.-F.} \bibnamefont{{Xu}}},
  \bibinfo{author}{\bibfnamefont{S.-K.} \bibnamefont{{Wang}}},
  \bibinfo{author}{\bibfnamefont{D.}~\bibnamefont{{Yang}}},
  \bibinfo{author}{\bibfnamefont{Y.-F.} \bibnamefont{{Hu}}},
  \bibinfo{author}{\bibfnamefont{S.}~\bibnamefont{{Jiang}}},
  \bibnamefont{et~al.}, \bibinfo{journal}{Nature Photonics}
  \textbf{\bibinfo{volume}{4}}, \bibinfo{pages}{376} (\bibinfo{year}{2010}).

\bibitem[{\citenamefont{Yin et~al.}(2012)\citenamefont{Yin, Ren, Lu, Cao, Yong,
  Wu, Liu, Liao, Zhou, Jiang et~al.}}]{Yin:Teleportation:2012}
\bibinfo{author}{\bibfnamefont{J.}~\bibnamefont{Yin}},
  \bibinfo{author}{\bibfnamefont{J.-G.} \bibnamefont{Ren}},
  \bibinfo{author}{\bibfnamefont{H.}~\bibnamefont{Lu}},
  \bibinfo{author}{\bibfnamefont{Y.}~\bibnamefont{Cao}},
  \bibinfo{author}{\bibfnamefont{H.-L.} \bibnamefont{Yong}},
  \bibinfo{author}{\bibfnamefont{Y.-P.} \bibnamefont{Wu}},
  \bibinfo{author}{\bibfnamefont{C.}~\bibnamefont{Liu}},
  \bibinfo{author}{\bibfnamefont{S.-K.} \bibnamefont{Liao}},
  \bibinfo{author}{\bibfnamefont{F.}~\bibnamefont{Zhou}},
  \bibinfo{author}{\bibfnamefont{Y.}~\bibnamefont{Jiang}},
  \bibnamefont{et~al.}, \bibinfo{journal}{Nature}
  \textbf{\bibinfo{volume}{488}}, \bibinfo{pages}{185} (\bibinfo{year}{2012}),
  ISSN \bibinfo{issn}{0028-0836}.

\bibitem[{\citenamefont{Ma et~al.}(2012)\citenamefont{Ma, Herbst, Scheidl,
  Wang, Kropatschek, Naylor, Wittmann, Mech, Kofler, Anisimova
  et~al.}}]{XSMa:Teleportation:2012}
\bibinfo{author}{\bibfnamefont{X.-S.} \bibnamefont{Ma}},
  \bibinfo{author}{\bibfnamefont{T.}~\bibnamefont{Herbst}},
  \bibinfo{author}{\bibfnamefont{T.}~\bibnamefont{Scheidl}},
  \bibinfo{author}{\bibfnamefont{D.}~\bibnamefont{Wang}},
  \bibinfo{author}{\bibfnamefont{S.}~\bibnamefont{Kropatschek}},
  \bibinfo{author}{\bibfnamefont{W.}~\bibnamefont{Naylor}},
  \bibinfo{author}{\bibfnamefont{B.}~\bibnamefont{Wittmann}},
  \bibinfo{author}{\bibfnamefont{A.}~\bibnamefont{Mech}},
  \bibinfo{author}{\bibfnamefont{J.}~\bibnamefont{Kofler}},
  \bibinfo{author}{\bibfnamefont{E.}~\bibnamefont{Anisimova}},
  \bibnamefont{et~al.}, \bibinfo{journal}{Nature}
  \textbf{\bibinfo{volume}{489}}, \bibinfo{pages}{269} (\bibinfo{year}{2012}).

\bibitem[{\citenamefont{Wang et~al.}(2013)\citenamefont{Wang, Yang, Liao,
  Zhang, Shen, Hu, Wu, Yang, Jiang, Tang et~al.}}]{Wang:Direct:2013}
\bibinfo{author}{\bibfnamefont{J.-Y.} \bibnamefont{Wang}},
  \bibinfo{author}{\bibfnamefont{B.}~\bibnamefont{Yang}},
  \bibinfo{author}{\bibfnamefont{S.-K.} \bibnamefont{Liao}},
  \bibinfo{author}{\bibfnamefont{L.}~\bibnamefont{Zhang}},
  \bibinfo{author}{\bibfnamefont{Q.}~\bibnamefont{Shen}},
  \bibinfo{author}{\bibfnamefont{X.-F.} \bibnamefont{Hu}},
  \bibinfo{author}{\bibfnamefont{J.-C.} \bibnamefont{Wu}},
  \bibinfo{author}{\bibfnamefont{S.-J.} \bibnamefont{Yang}},
  \bibinfo{author}{\bibfnamefont{H.}~\bibnamefont{Jiang}},
  \bibinfo{author}{\bibfnamefont{Y.-L.} \bibnamefont{Tang}},
  \bibnamefont{et~al.}, \bibinfo{journal}{Nature Photonics}
  \textbf{\bibinfo{volume}{7}}, \bibinfo{pages}{387} (\bibinfo{year}{2013}),
  ISSN \bibinfo{issn}{1749-4885}.

\bibitem[{\citenamefont{Nauerth et~al.}(2013)\citenamefont{Nauerth, Moll, Rau,
  Fuchs, Horwath, Frick, and Weinfurter}}]{Nauerth:AirQKD:2013}
\bibinfo{author}{\bibfnamefont{S.}~\bibnamefont{Nauerth}},
  \bibinfo{author}{\bibfnamefont{F.}~\bibnamefont{Moll}},
  \bibinfo{author}{\bibfnamefont{M.}~\bibnamefont{Rau}},
  \bibinfo{author}{\bibfnamefont{C.}~\bibnamefont{Fuchs}},
  \bibinfo{author}{\bibfnamefont{J.}~\bibnamefont{Horwath}},
  \bibinfo{author}{\bibfnamefont{S.}~\bibnamefont{Frick}}, \bibnamefont{and}
  \bibinfo{author}{\bibfnamefont{H.}~\bibnamefont{Weinfurter}},
  \bibinfo{journal}{Nature Photonics} \textbf{\bibinfo{volume}{7}},
  \bibinfo{pages}{382} (\bibinfo{year}{2013}), ISSN \bibinfo{issn}{1749-4885}.

\bibitem[{\citenamefont{Yin et~al.}(2013)\citenamefont{Yin, Cao, Liu, Pan,
  Wang, Yang, Zhang, Yang, Chen, Peng et~al.}}]{Yin:single:2013}
\bibinfo{author}{\bibfnamefont{J.}~\bibnamefont{Yin}},
  \bibinfo{author}{\bibfnamefont{Y.}~\bibnamefont{Cao}},
  \bibinfo{author}{\bibfnamefont{S.-B.} \bibnamefont{Liu}},
  \bibinfo{author}{\bibfnamefont{G.-S.} \bibnamefont{Pan}},
  \bibinfo{author}{\bibfnamefont{J.-H.} \bibnamefont{Wang}},
  \bibinfo{author}{\bibfnamefont{T.}~\bibnamefont{Yang}},
  \bibinfo{author}{\bibfnamefont{Z.-P.} \bibnamefont{Zhang}},
  \bibinfo{author}{\bibfnamefont{F.-M.} \bibnamefont{Yang}},
  \bibinfo{author}{\bibfnamefont{Y.-A.} \bibnamefont{Chen}},
  \bibinfo{author}{\bibfnamefont{C.-Z.} \bibnamefont{Peng}},
  \bibnamefont{et~al.}, \bibinfo{journal}{Optics express}
  \textbf{\bibinfo{volume}{21}}, \bibinfo{pages}{20032} (\bibinfo{year}{2013}).

\bibitem[{\citenamefont{Vallone et~al.}(2015)\citenamefont{Vallone, Bacco,
  Dequal, Gaiarin, Luceri, Bianco, and Villoresi}}]{Vallone:Single:2015}
\bibinfo{author}{\bibfnamefont{G.}~\bibnamefont{Vallone}},
  \bibinfo{author}{\bibfnamefont{D.}~\bibnamefont{Bacco}},
  \bibinfo{author}{\bibfnamefont{D.}~\bibnamefont{Dequal}},
  \bibinfo{author}{\bibfnamefont{S.}~\bibnamefont{Gaiarin}},
  \bibinfo{author}{\bibfnamefont{V.}~\bibnamefont{Luceri}},
  \bibinfo{author}{\bibfnamefont{G.}~\bibnamefont{Bianco}}, \bibnamefont{and}
  \bibinfo{author}{\bibfnamefont{P.}~\bibnamefont{Villoresi}},
  \bibinfo{journal}{Phys. Rev. Lett.} \textbf{\bibinfo{volume}{115}},
  \bibinfo{pages}{040502} (\bibinfo{year}{2015}).

\bibitem[{\citenamefont{Xin}(2011)}]{Xin:Chinese:2011}
\bibinfo{author}{\bibfnamefont{H.}~\bibnamefont{Xin}},
  \bibinfo{journal}{Science} \textbf{\bibinfo{volume}{332}},
  \bibinfo{pages}{904} (\bibinfo{year}{2011}).

\bibitem[{\citenamefont{Fritz}(1996)}]{Fritz:Revisit:1996}
\bibinfo{author}{\bibfnamefont{L.~W.} \bibnamefont{Fritz}},
  \bibinfo{journal}{International Archives of Photogrammetry and Remote
  Sensing} \textbf{\bibinfo{volume}{31}}, \bibinfo{pages}{273}
  (\bibinfo{year}{1996}).

\bibitem[{\citenamefont{Pratt et~al.}(1999)\citenamefont{Pratt, Raines,
  Fossa~Jr, Temple et~al.}}]{Pratt:IRIDIUM:1999}
\bibinfo{author}{\bibfnamefont{S.~R.} \bibnamefont{Pratt}},
  \bibinfo{author}{\bibfnamefont{R.}~\bibnamefont{Raines}},
  \bibinfo{author}{\bibfnamefont{C.~E.} \bibnamefont{Fossa~Jr}},
  \bibinfo{author}{\bibfnamefont{M.}~\bibnamefont{Temple}},
  \bibnamefont{et~al.}, \bibinfo{journal}{Communications Surveys, IEEE}
  \textbf{\bibinfo{volume}{2}}, \bibinfo{pages}{2} (\bibinfo{year}{1999}).

\bibitem[{\citenamefont{Shentu et~al.}(2013)\citenamefont{Shentu, Pelc, Wang,
  Sun, Zheng, Fejer, Zhang, and Pan}}]{Shentu:Detector:2013}
\bibinfo{author}{\bibfnamefont{G.-L.} \bibnamefont{Shentu}},
  \bibinfo{author}{\bibfnamefont{J.~S.} \bibnamefont{Pelc}},
  \bibinfo{author}{\bibfnamefont{X.-D.} \bibnamefont{Wang}},
  \bibinfo{author}{\bibfnamefont{Q.-C.} \bibnamefont{Sun}},
  \bibinfo{author}{\bibfnamefont{M.-Y.} \bibnamefont{Zheng}},
  \bibinfo{author}{\bibfnamefont{M.}~\bibnamefont{Fejer}},
  \bibinfo{author}{\bibfnamefont{Q.}~\bibnamefont{Zhang}}, \bibnamefont{and}
  \bibinfo{author}{\bibfnamefont{J.-W.} \bibnamefont{Pan}},
  \bibinfo{journal}{Optics express} \textbf{\bibinfo{volume}{21}},
  \bibinfo{pages}{13986} (\bibinfo{year}{2013}).

\bibitem[{\citenamefont{Gilmore}(2002)}]{Gilmore:spacecraft_thermal:2002}
\bibinfo{author}{\bibfnamefont{D.~G.} \bibnamefont{Gilmore}},
  \emph{\bibinfo{title}{Spacecraft thermal control handbook: fundamental
  technologies}}, vol.~\bibinfo{volume}{1} (\bibinfo{publisher}{AIAA},
  \bibinfo{year}{2002}).

\bibitem[{\citenamefont{Pfennigbauer et~al.}(2003)\citenamefont{Pfennigbauer,
  Leeb, Aspelmeyer, Jennewein, and
  Zeilinger}}]{Pfennigbauer:Intersatellite:2003}
\bibinfo{author}{\bibfnamefont{M.}~\bibnamefont{Pfennigbauer}},
  \bibinfo{author}{\bibfnamefont{W.}~\bibnamefont{Leeb}},
  \bibinfo{author}{\bibfnamefont{M.}~\bibnamefont{Aspelmeyer}},
  \bibinfo{author}{\bibfnamefont{T.}~\bibnamefont{Jennewein}},
  \bibnamefont{and}
  \bibinfo{author}{\bibfnamefont{A.}~\bibnamefont{Zeilinger}},
  \emph{\bibinfo{title}{Free-space optical quantum key distribution using
  intersatellite links}} (\bibinfo{publisher}{CNES - Intersatellite Link
  Workshop}, \bibinfo{year}{2003}).

\bibitem[{\citenamefont{Tomaello et~al.}(2011)\citenamefont{Tomaello,
  Dall'Arche, Naletto, and Villoresi}}]{Tomaello:intersatellite:2011}
\bibinfo{author}{\bibfnamefont{A.}~\bibnamefont{Tomaello}},
  \bibinfo{author}{\bibfnamefont{A.}~\bibnamefont{Dall'Arche}},
  \bibinfo{author}{\bibfnamefont{G.}~\bibnamefont{Naletto}}, \bibnamefont{and}
  \bibinfo{author}{\bibfnamefont{P.}~\bibnamefont{Villoresi}}, in
  \emph{\bibinfo{booktitle}{SPIE Optical Engineering+ Applications}}
  (\bibinfo{organization}{International Society for Optics and Photonics},
  \bibinfo{year}{2011}), pp. \bibinfo{pages}{816309--816309}.

\bibitem[{\citenamefont{Buttler et~al.}(2000)\citenamefont{Buttler, Hughes,
  Lamoreaux, Morgan, Nordholt, and Peterson}}]{Buttler:Daylight:2000}
\bibinfo{author}{\bibfnamefont{W.~T.} \bibnamefont{Buttler}},
  \bibinfo{author}{\bibfnamefont{R.~J.} \bibnamefont{Hughes}},
  \bibinfo{author}{\bibfnamefont{S.~K.} \bibnamefont{Lamoreaux}},
  \bibinfo{author}{\bibfnamefont{G.~L.} \bibnamefont{Morgan}},
  \bibinfo{author}{\bibfnamefont{J.~E.} \bibnamefont{Nordholt}},
  \bibnamefont{and} \bibinfo{author}{\bibfnamefont{C.~G.}
  \bibnamefont{Peterson}}, \bibinfo{journal}{Physical Review Letters}
  \textbf{\bibinfo{volume}{84}}, \bibinfo{pages}{5652} (\bibinfo{year}{2000}).

\bibitem[{\citenamefont{Hughes et~al.}(2002)\citenamefont{Hughes, Nordholt,
  Derkacs, and Peterson}}]{Hughes2002:10km}
\bibinfo{author}{\bibfnamefont{R.~J.} \bibnamefont{Hughes}},
  \bibinfo{author}{\bibfnamefont{J.~E.} \bibnamefont{Nordholt}},
  \bibinfo{author}{\bibfnamefont{D.}~\bibnamefont{Derkacs}}, \bibnamefont{and}
  \bibinfo{author}{\bibfnamefont{C.~G.} \bibnamefont{Peterson}},
  \bibinfo{journal}{New journal of physics} \textbf{\bibinfo{volume}{4}},
  \bibinfo{pages}{43} (\bibinfo{year}{2002}).

\bibitem[{\citenamefont{H{\"o}ckel et~al.}(2009)\citenamefont{H{\"o}ckel, Koch,
  Martin, and Benson}}]{Hockel:Ultranarrow:2009}
\bibinfo{author}{\bibfnamefont{D.}~\bibnamefont{H{\"o}ckel}},
  \bibinfo{author}{\bibfnamefont{L.}~\bibnamefont{Koch}},
  \bibinfo{author}{\bibfnamefont{E.}~\bibnamefont{Martin}}, \bibnamefont{and}
  \bibinfo{author}{\bibfnamefont{O.}~\bibnamefont{Benson}},
  \bibinfo{journal}{Optics letters} \textbf{\bibinfo{volume}{34}},
  \bibinfo{pages}{3169} (\bibinfo{year}{2009}).

\bibitem[{\citenamefont{Restelli et~al.}(2010)\citenamefont{Restelli, Bienfang,
  Clark, Rech, Labanca, Ghioni, and Cova}}]{Restelli:Timing:Daylight:2010}
\bibinfo{author}{\bibfnamefont{A.}~\bibnamefont{Restelli}},
  \bibinfo{author}{\bibfnamefont{J.~C.} \bibnamefont{Bienfang}},
  \bibinfo{author}{\bibfnamefont{C.~W.} \bibnamefont{Clark}},
  \bibinfo{author}{\bibfnamefont{I.}~\bibnamefont{Rech}},
  \bibinfo{author}{\bibfnamefont{I.}~\bibnamefont{Labanca}},
  \bibinfo{author}{\bibfnamefont{M.}~\bibnamefont{Ghioni}}, \bibnamefont{and}
  \bibinfo{author}{\bibfnamefont{S.}~\bibnamefont{Cova}},
  \bibinfo{journal}{Selected Topics in Quantum Electronics, IEEE Journal of}
  \textbf{\bibinfo{volume}{16}}, \bibinfo{pages}{1084} (\bibinfo{year}{2010}).

\bibitem[{\citenamefont{Shan et~al.}(2006)\citenamefont{Shan, Sun, Luo, Tan,
  and Zhan}}]{Shan:Rbvapor:Daylight:2006}
\bibinfo{author}{\bibfnamefont{X.}~\bibnamefont{Shan}},
  \bibinfo{author}{\bibfnamefont{X.}~\bibnamefont{Sun}},
  \bibinfo{author}{\bibfnamefont{J.}~\bibnamefont{Luo}},
  \bibinfo{author}{\bibfnamefont{Z.}~\bibnamefont{Tan}}, \bibnamefont{and}
  \bibinfo{author}{\bibfnamefont{M.}~\bibnamefont{Zhan}},
  \bibinfo{journal}{Applied physics letters} \textbf{\bibinfo{volume}{89}},
  \bibinfo{pages}{191121} (\bibinfo{year}{2006}).

\bibitem[{\citenamefont{Rogers et~al.}(2006)\citenamefont{Rogers, Bienfang,
  Mink, Hershman, Nakassis, Tang, Ma, Su, Williams, and
  Clark}}]{Rogers:Halpha:Daylight:2006}
\bibinfo{author}{\bibfnamefont{D.}~\bibnamefont{Rogers}},
  \bibinfo{author}{\bibfnamefont{J.}~\bibnamefont{Bienfang}},
  \bibinfo{author}{\bibfnamefont{A.}~\bibnamefont{Mink}},
  \bibinfo{author}{\bibfnamefont{B.~J.} \bibnamefont{Hershman}},
  \bibinfo{author}{\bibfnamefont{A.}~\bibnamefont{Nakassis}},
  \bibinfo{author}{\bibfnamefont{X.}~\bibnamefont{Tang}},
  \bibinfo{author}{\bibfnamefont{L.}~\bibnamefont{Ma}},
  \bibinfo{author}{\bibfnamefont{D.~H.} \bibnamefont{Su}},
  \bibinfo{author}{\bibfnamefont{C.~J.} \bibnamefont{Williams}},
  \bibnamefont{and} \bibinfo{author}{\bibfnamefont{C.~W.} \bibnamefont{Clark}},
  in \emph{\bibinfo{booktitle}{SPIE Optics+ Photonics}}
  (\bibinfo{organization}{International Society for Optics and Photonics},
  \bibinfo{year}{2006}), pp. \bibinfo{pages}{630417--630417}.

\bibitem[{\citenamefont{Peloso et~al.}(2009)\citenamefont{Peloso, Gerhardt, Ho,
  Lamas-Linares, and Kurtsiefer}}]{Peloso:Entanglement:Daylight:2009}
\bibinfo{author}{\bibfnamefont{M.~P.} \bibnamefont{Peloso}},
  \bibinfo{author}{\bibfnamefont{I.}~\bibnamefont{Gerhardt}},
  \bibinfo{author}{\bibfnamefont{C.}~\bibnamefont{Ho}},
  \bibinfo{author}{\bibfnamefont{A.}~\bibnamefont{Lamas-Linares}},
  \bibnamefont{and}
  \bibinfo{author}{\bibfnamefont{C.}~\bibnamefont{Kurtsiefer}},
  \bibinfo{journal}{New Journal of Physics} \textbf{\bibinfo{volume}{11}},
  \bibinfo{pages}{045007} (\bibinfo{year}{2009}).

\bibitem[{\citenamefont{Miao et~al.}(2005)\citenamefont{Miao, Han, Gong, Zhang,
  Diao, and Guo}}]{Miao:QKDNoise:2005}
\bibinfo{author}{\bibfnamefont{E.-l.} \bibnamefont{Miao}},
  \bibinfo{author}{\bibfnamefont{Z.-f.} \bibnamefont{Han}},
  \bibinfo{author}{\bibfnamefont{S.-s.} \bibnamefont{Gong}},
  \bibinfo{author}{\bibfnamefont{T.}~\bibnamefont{Zhang}},
  \bibinfo{author}{\bibfnamefont{D.-s.} \bibnamefont{Diao}}, \bibnamefont{and}
  \bibinfo{author}{\bibfnamefont{G.-c.} \bibnamefont{Guo}},
  \bibinfo{journal}{New Journal of Physics} \textbf{\bibinfo{volume}{7}},
  \bibinfo{pages}{215} (\bibinfo{year}{2005}).

\bibitem[{\citenamefont{Ren et~al.}(2009)\citenamefont{Ren, Yang, Yi, Zhou,
  Chen, Peng, and Pan}}]{Ren:Teleportation:2009}
\bibinfo{author}{\bibfnamefont{J.-G.} \bibnamefont{Ren}},
  \bibinfo{author}{\bibfnamefont{B.}~\bibnamefont{Yang}},
  \bibinfo{author}{\bibfnamefont{Z.-H.} \bibnamefont{Yi}},
  \bibinfo{author}{\bibfnamefont{F.}~\bibnamefont{Zhou}},
  \bibinfo{author}{\bibfnamefont{K.}~\bibnamefont{Chen}},
  \bibinfo{author}{\bibfnamefont{C.-Z.} \bibnamefont{Peng}}, \bibnamefont{and}
  \bibinfo{author}{\bibfnamefont{J.-W.} \bibnamefont{Pan}},
  \bibinfo{journal}{Chinese Physics B} \textbf{\bibinfo{volume}{18}},
  \bibinfo{pages}{3605} (\bibinfo{year}{2009}).

\bibitem[{\citenamefont{Hwang}(2003)}]{Hwang:Decoy:2003}
\bibinfo{author}{\bibfnamefont{W.-Y.} \bibnamefont{Hwang}},
  \bibinfo{journal}{Phys.~Rev.~Lett.~} \textbf{\bibinfo{volume}{91}},
  \bibinfo{pages}{057901} (\bibinfo{year}{2003}).

\bibitem[{\citenamefont{Wang}(2005)}]{Wang:Decoy:2005}
\bibinfo{author}{\bibfnamefont{X.-B.} \bibnamefont{Wang}},
  \bibinfo{journal}{Phys.~Rev.~Lett.~} \textbf{\bibinfo{volume}{94}},
  \bibinfo{pages}{230503} (\bibinfo{year}{2005}).

\bibitem[{\citenamefont{Lo et~al.}(2005)\citenamefont{Lo, Ma, and
  Chen}}]{Lo:Decoy:2005}
\bibinfo{author}{\bibfnamefont{H.-K.} \bibnamefont{Lo}},
  \bibinfo{author}{\bibfnamefont{X.}~\bibnamefont{Ma}}, \bibnamefont{and}
  \bibinfo{author}{\bibfnamefont{K.}~\bibnamefont{Chen}},
  \bibinfo{journal}{Phys.~Rev.~Lett.~} \textbf{\bibinfo{volume}{94}},
  \bibinfo{pages}{230504} (\bibinfo{year}{2005}).

\bibitem[{\citenamefont{Fung et~al.}(2010)\citenamefont{Fung, Ma, and
  Chau}}]{Fung:Finite:2010}
\bibinfo{author}{\bibfnamefont{C.-H.~F.} \bibnamefont{Fung}},
  \bibinfo{author}{\bibfnamefont{X.}~\bibnamefont{Ma}}, \bibnamefont{and}
  \bibinfo{author}{\bibfnamefont{H.~F.} \bibnamefont{Chau}},
  \bibinfo{journal}{Phys. Rev. A} \textbf{\bibinfo{volume}{81}},
  \bibinfo{pages}{012318} (\bibinfo{year}{2010}).

\bibitem[{\citenamefont{Huttner et~al.}(1995)\citenamefont{Huttner, Imoto,
  Gisin, and Mor}}]{HIGM:PNS:1995}
\bibinfo{author}{\bibfnamefont{B.}~\bibnamefont{Huttner}},
  \bibinfo{author}{\bibfnamefont{N.}~\bibnamefont{Imoto}},
  \bibinfo{author}{\bibfnamefont{N.}~\bibnamefont{Gisin}}, \bibnamefont{and}
  \bibinfo{author}{\bibfnamefont{T.}~\bibnamefont{Mor}},
  \bibinfo{journal}{Phys.~Rev.~A} \textbf{\bibinfo{volume}{51}},
  \bibinfo{pages}{1863} (\bibinfo{year}{1995}).

\bibitem[{\citenamefont{Brassard et~al.}(2000)\citenamefont{Brassard,
  L\"utkenhaus, Mor, and Sanders}}]{BLMS:PNS:2000}
\bibinfo{author}{\bibfnamefont{G.}~\bibnamefont{Brassard}},
  \bibinfo{author}{\bibfnamefont{N.}~\bibnamefont{L\"utkenhaus}},
  \bibinfo{author}{\bibfnamefont{T.}~\bibnamefont{Mor}}, \bibnamefont{and}
  \bibinfo{author}{\bibfnamefont{B.~C.} \bibnamefont{Sanders}},
  \bibinfo{journal}{Phys.~Rev.~Lett.~} \textbf{\bibinfo{volume}{85}},
  \bibinfo{pages}{1330} (\bibinfo{year}{2000}).

\bibitem[{\citenamefont{Lo and Preskill}(2007)}]{LoPreskill:NonRan:2007}
\bibinfo{author}{\bibfnamefont{H.-K.} \bibnamefont{Lo}} \bibnamefont{and}
  \bibinfo{author}{\bibfnamefont{J.}~\bibnamefont{Preskill}},
  \bibinfo{journal}{Quantum Inf.~Comput.} \textbf{\bibinfo{volume}{7}},
  \bibinfo{pages}{0431} (\bibinfo{year}{2007}).

\bibitem[{\citenamefont{Lo et~al.}(2012)\citenamefont{Lo, Curty, and
  Qi}}]{Lo:MIQKD:2012}
\bibinfo{author}{\bibfnamefont{H.-K.} \bibnamefont{Lo}},
  \bibinfo{author}{\bibfnamefont{M.}~\bibnamefont{Curty}}, \bibnamefont{and}
  \bibinfo{author}{\bibfnamefont{B.}~\bibnamefont{Qi}}, \bibinfo{journal}{Phys.
  Rev. Lett.} \textbf{\bibinfo{volume}{108}}, \bibinfo{pages}{130503}
  (\bibinfo{year}{2012}).

\bibitem[{\citenamefont{Bennett et~al.}(1993)\citenamefont{Bennett, Brassard,
  Cr\'{e}peau, Jozsa, Peres, and Wootters}}]{BBCJPW_93}
\bibinfo{author}{\bibfnamefont{C.~H.} \bibnamefont{Bennett}},
  \bibinfo{author}{\bibfnamefont{G.}~\bibnamefont{Brassard}},
  \bibinfo{author}{\bibfnamefont{C.}~\bibnamefont{Cr\'{e}peau}},
  \bibinfo{author}{\bibfnamefont{R.}~\bibnamefont{Jozsa}},
  \bibinfo{author}{\bibfnamefont{A.}~\bibnamefont{Peres}}, \bibnamefont{and}
  \bibinfo{author}{\bibfnamefont{W.~K.} \bibnamefont{Wootters}},
  \bibinfo{journal}{Phys.~Rev.~Lett.~} \textbf{\bibinfo{volume}{70}},
  \bibinfo{pages}{1895} (\bibinfo{year}{1993}).

\bibitem[{\citenamefont{Briegel et~al.}(1998)\citenamefont{Briegel, D{\"u}r,
  Cirac, and Zoller}}]{Briegel1998:repeater}
\bibinfo{author}{\bibfnamefont{H.-J.} \bibnamefont{Briegel}},
  \bibinfo{author}{\bibfnamefont{W.}~\bibnamefont{D{\"u}r}},
  \bibinfo{author}{\bibfnamefont{J.~I.} \bibnamefont{Cirac}}, \bibnamefont{and}
  \bibinfo{author}{\bibfnamefont{P.}~\bibnamefont{Zoller}},
  \bibinfo{journal}{Physical Review Letters} \textbf{\bibinfo{volume}{81}},
  \bibinfo{pages}{5932} (\bibinfo{year}{1998}).

\bibitem[{\citenamefont{Giovannetti et~al.}(2004)\citenamefont{Giovannetti,
  Lloyd, and Maccone}}]{Giovannetti:QEM:2004}
\bibinfo{author}{\bibfnamefont{V.}~\bibnamefont{Giovannetti}},
  \bibinfo{author}{\bibfnamefont{S.}~\bibnamefont{Lloyd}}, \bibnamefont{and}
  \bibinfo{author}{\bibfnamefont{L.}~\bibnamefont{Maccone}},
  \bibinfo{journal}{Science} \textbf{\bibinfo{volume}{306}},
  \bibinfo{pages}{1330} (\bibinfo{year}{2004}).

\bibitem[{\citenamefont{Yurke}(1985)}]{Yurke:Homodyne:1985}
\bibinfo{author}{\bibfnamefont{B.}~\bibnamefont{Yurke}},
  \bibinfo{journal}{Physical Review A} \textbf{\bibinfo{volume}{32}},
  \bibinfo{pages}{311} (\bibinfo{year}{1985}).

\bibitem[{\citenamefont{Komar et~al.}(2014)\citenamefont{Komar, Kessler,
  Bishof, Jiang, S{\o}rensen, Ye, and Lukin}}]{Komar:clock:2014}
\bibinfo{author}{\bibfnamefont{P.}~\bibnamefont{Komar}},
  \bibinfo{author}{\bibfnamefont{E.~M.} \bibnamefont{Kessler}},
  \bibinfo{author}{\bibfnamefont{M.}~\bibnamefont{Bishof}},
  \bibinfo{author}{\bibfnamefont{L.}~\bibnamefont{Jiang}},
  \bibinfo{author}{\bibfnamefont{A.~S.} \bibnamefont{S{\o}rensen}},
  \bibinfo{author}{\bibfnamefont{J.}~\bibnamefont{Ye}}, \bibnamefont{and}
  \bibinfo{author}{\bibfnamefont{M.~D.} \bibnamefont{Lukin}},
  \bibinfo{journal}{Nature Physics}  (\bibinfo{year}{2014}).

\bibitem[{\citenamefont{Thew et~al.}(2006)\citenamefont{Thew, Tanzilli,
  Krainer, Zeller, Rochas, Rech, Cova, Zbinden, and Gisin}}]{Gisin_up2_06}
\bibinfo{author}{\bibfnamefont{R.~T.} \bibnamefont{Thew}},
  \bibinfo{author}{\bibfnamefont{S.}~\bibnamefont{Tanzilli}},
  \bibinfo{author}{\bibfnamefont{L.}~\bibnamefont{Krainer}},
  \bibinfo{author}{\bibfnamefont{S.~C.} \bibnamefont{Zeller}},
  \bibinfo{author}{\bibfnamefont{A.}~\bibnamefont{Rochas}},
  \bibinfo{author}{\bibfnamefont{I.}~\bibnamefont{Rech}},
  \bibinfo{author}{\bibfnamefont{S.}~\bibnamefont{Cova}},
  \bibinfo{author}{\bibfnamefont{H.}~\bibnamefont{Zbinden}}, \bibnamefont{and}
  \bibinfo{author}{\bibfnamefont{N.}~\bibnamefont{Gisin}},
  \bibinfo{journal}{New Journal of Physics} \textbf{\bibinfo{volume}{8}},
  \bibinfo{pages}{32} (\bibinfo{year}{2006}).

\end{thebibliography}
\end{document}